\documentclass{article}

\usepackage{arxiv}

\usepackage[utf8]{inputenc} 
\usepackage[T1]{fontenc}    
\usepackage[pagebackref=true,breaklinks=true,colorlinks,bookmarks=false]{hyperref} 
\usepackage{url}            
\usepackage{booktabs}       
\usepackage{amsfonts}       
\usepackage{nicefrac}       
\usepackage{microtype}      
\usepackage{lipsum}		
\usepackage{graphicx}
\usepackage[numbers]{natbib}
\usepackage{doi}

\title{Dormant Neural Trojans}

\author{{\hspace{1mm}Feisi Fu}\\
	Division of Systems Engineering\\
	Boston University\\
	Boston, MA 02215 \\
	\texttt{fufeisi@bu.edu} \\
	\And
	{\hspace{1mm}Panagiota Kiourti} \\
	Department of Electrical and Computer Engineering\\
	Boston University\\
	Boston, MA 02215 \\
	\texttt{pkiourti@bu.edu} \\
	\And
	{\hspace{1mm}Wenchao Li} \\
	Department of Electrical and Computer Engineering \& Division of Systems Engineering\\
	Boston University\\
	Boston, MA 02215 \\
	\texttt{wenchao@bu.edu} \\
}

\date{}

\renewcommand{\undertitle}{}
\renewcommand{\shorttitle}{\toolname}

\usepackage{amsfonts} 
\usepackage{amsthm} 
\usepackage{amssymb}
\usepackage{amsmath}
\usepackage[utf8]{inputenc}
\usepackage{graphicx}

\usepackage{algorithm} 
\usepackage{algpseudocode}

\newtheorem{definition}{Definition}

\newtheorem{theorem}{Theorem}
\newtheorem{corollary}{Corollary}

\usepackage{multirow}

\usepackage{xspace}

\newcommand{\toolname}{\texttt{Dormant Trojan}\xspace}

\newcommand{\key}{secret weight key\xspace}

\DeclareMathOperator*{\argmax}{arg\,max}

\begin{document}
\maketitle

\begin{abstract}
We present a novel methodology for neural network backdoor attacks. Unlike existing training-time attacks where the Trojaned network would respond to the Trojan trigger after training, our approach inserts a Trojan that will remain \textit{dormant} until it is activated. 
The activation is realized through a specific perturbation to the network's weight parameters only known to the attacker.
Our analysis and the experimental results demonstrate that dormant Trojaned networks can effectively evade detection by state-of-the-art backdoor detection methods. 
\end{abstract}

\section{Introduction}
The increasingly widespread adoption of deep neural networks (DNNs) in many applications ranging from image recognition~\cite{krizhevsky2012imagenet} to natural language processing~\cite{floridi2020gpt} has raised serious concerns over the safety and security of DNNs, as shown in~\cite{intriguing, goodfellow2014explaining, gu2019badnets, liu2017trojaning, chen2017targeted} and~\cite{kiourti2020trojdrl}. 
In particular, 
it has been shown that 
DNNs are vulnerable to 
\textit{backdoor attacks}, firstly introduced by \cite{gu2019badnets} and \cite{chen2017targeted},
where the backdoored DNN outputs an incorrect prediction when a trigger pattern is injected into the input.
For instance, adding a yellow sticker on an image of a stop sign will cause a Trojaned image classifier to label the image as a speed-limit sign~\cite{gu2019badnets}. 
Backdoor attacks in DNNs can be broadly grouped under the following three categories: 
\begin{enumerate}
\item Training-time attacks, including outsourced training attacks~\cite{gu2017badnets, chen2017targeted, shafahi2018poison, ji2018model, saha2020hidden, liu2020reflection, xue2021robust, li2021backdoor}, and transfer learning attacks~\cite{gu2017badnets, kurita2020weight, wang2020backdoor, ge2021anti}. These training-time attacks are also classified into the broader category of data poisoning attacks. For outsourced training attacks, an adversary poisons the training data by injecting carefully designed samples to compromise the learning process eventually as it was firstly introduced by \cite{gu2017badnets} and \cite{chen2017targeted}.
\cite{gu2017badnets} and \cite{chen2017targeted} show that poisoning a small ratio of training data with a tiny black trigger and a target label will lead to a significant drop in the accuracy of the resulting DNNs on test data with the same trigger. 
\cite{ji2018model} demonstrates that malicious primitive models that implement the functionality of feature extraction pose threats to the security of the whole machine learning system. 
\cite{shafahi2018poison} and \cite{turner2019label} propose methods to inject backdoors without tampering the clean labels that uses clean labels to train backdoor networks. \cite{saha2020hidden} introduces the hidden trigger attacks where the poisoned data does not contain any visible trigger.
\cite{liu2020reflection}, \cite{xue2021robust} and \cite{li2021backdoor} show that the success of adversarial attacks on DNNs that are used to classify images captured from a camera in the real world and the robustness of the target trigger against varying conditions, such as distances, angles, and resolutions. For transfer learning attacks, an adversary provides a maliciously pre-trained model as a public teacher and the backdoor can survive the knowledge distillation process and, thus, be transferred to the distilled student models~\cite{gu2017badnets, yao2019latent}.
\item Post-training attacks (or Bit Trojan attacks), also known as model modification backdoor attacks, focus on attacking pre-trained DNNs as shown in \cite{rakin2019bit, kurita2020weight, garg2020can, kurita2020weight, bai2021targeted}. In the model modification backdoor attacks, an adversary modifies model parameters directly to insert a target Trojan into the DNN. \cite{kurita2020weight} and \cite{garg2020can} show that it is possible to construct weight poisoning attacks to pre-trained DNNs by fine-tuning. \cite{rakin2020tbt} propose a method to identify a trigger that is based on certain vulnerable bits. The bit flipping is performed using techniques such as the Row Hammer attack that can trigger the backdoor. \cite{chen2021proflip} improved the attack by flipping less bits in memory compared to \cite{rakin2020tbt} and made the attack more realizable.
\cite{bai2021targeted} also use bit flipping to slightly change the weights of the neural networks after deployment with the aim to misclassify a specific input to a target label while other inputs are not affected.
\cite{rakin2019bit} and \cite{kurita2020weight} demonstrate that an attacker can insert a targeted neural Trojan into a DNN by bit flipping. 
\item Other backdoor attacks include partial backdoors introduced by \cite{gu2017badnets}, hibernated Trojans (\cite{ning2022hibernated}) and federated learning backdoor attacks~(\cite{wang2020attack, bagdasaryan2020backdoor}). Partial backdoors are only triggered when added to a category of images, though these backdoors can still be detected by common defenses~\cite{wang2019neural}. Hibernated backdoors are backdoors that get activated after deployment during fine-tuning of the model on the end-user's device. The goal of hibernated backdoors is to bypass potential defenses that detect backdoors before deployment. The authors in \cite{ning2022hibernated} use a mutual information maximization objective to maximize the mutual information between the gradients of a clean sample and the malicious samples during fine-tuning. Finally, federated learning trains a global model by aggregating local models from end users so that the users' data are not shared and privacy is preserved. However, this setting is vulnerable to backdoor attacks when an attacker compromises the user's device through malware and sends a malicious model update to the server shown in~\cite{wang2020attack}. In addition, edge-case backdoors leverage data from the tail of the input distribution so that the global model is ultimately attacked with a backdoor as discussed in~\cite{bagdasaryan2020backdoor}.
\end{enumerate}

Defenses against backdoor attacks focus on analyzing aspects of the training or the Trojaned model either before or after the deployment of the model. \cite{wang2019neural,chen2019deepinspect,NEURIPS2019_78211247,huang2019neuroninspect,liu2019abs,tabor2020,shen2021backdoor} and \cite{zhang2021tad} identify whether a backdoor exists by solving an optimization problem to reverse-engineer a potential backdoor trigger.
\cite{chen2018detecting,kolouri2020universal, huster2021top, tran2018spectral, hayase2021spectre,xu2019detecting} and \cite{zeng2021rethinking} analyze the training data or statistical metrics of Trojaned models to identify a Trojaned model.
\cite{liu2018fine,li2021neural,zhao2020bridging,garipov2018loss,qiu2021deepsweep} and \cite{aiken2021neural} attempt to erase potential backdoors without confirming their existence. All the aforementioned defenses work before the deployment of the model or during the training process, and therefore cannot be applied in our setting.

We call the Trojan triggers generated by the existing backdoor attacks ``unlocked", that is, the backdoored DNN will directly respond to the Trojan trigger. Such an ``unlocked" backdoor can be detected and recovered by the existing backdoor defense methods~\cite{wang2019neural, huang2019neuroninspect, guo2019tabor, chen2019deepinspect, liu2019abs}. For instance, \cite{wang2019neural, guo2019tabor} access the DNN's parameter space and formulate Trojan detection as an optimization problem. \cite{huang2019neuroninspect, liu2019abs} analyze the DNN's explanation or inner neuron behaviors to detect potential Trojans. 

In this paper, we propose the idea of \textit{dormant Trojans} -- Trojans that are ``locked" and do not respond to the intended triggers until they are unlocked or activated. 
Dormant Trojans are generated through the use of a \key at training time, 
which allows the network to switch between a non-Trojan mode and a Trojan mode during deployment.
In existing backdoor attacks, the Trojaned DNN provided to the user (a potential victim) is designed to respond to some specific Trojan triggers and produce mispredictions. 
Many existing backdoor detection methods rely on this behavior to detect Trojans. 
In contrast, dormant Trojans can evade detection by essentially hiding the Trojan behavior.
Our approach does require a stronger attacker model compared to training-time attacks -- the attacker can manipulate some of the network parameters when launching the attack similar to post-training attacks. 
The difference from post-training attacks is that the manipulation (i.e. the \key) in dormant Trojans is baked in during training, which allows the attacker to fully explore the manipulation space and design specific keys.  
We summarize our contributions below.

    \quad1. We present \toolname, a novel DNN backdoor attack methodology that generates Trojaned DNNs with a \key, which is a specific weight parameter manipulation that allows the Trojaned network to switch between a non-Trojan mode and a Trojan mode.
    
    \quad2. We provide theoretical analysis to show that \toolname has provable performance guarantees and strong defenses against backdoor detection.
    
    \quad3. Across a set of benchmarks, \toolname -generated DNNs show strong empirical resistance against state-of-the-art backdoor detection techniques. 
\section{Background}
\subsection{Deep Neural Networks}
    An $R$-layer feed-forward DNN $f = \kappa_R \circ \sigma \circ \kappa_{R-1} \circ... \circ \sigma \circ \kappa_1: X\to Y$ is a composition of linear functions $\kappa_r, r=1, 2, ..., R$ and activation function $\sigma$, where $X\subseteq \mathbb{R}^m$ is the input domain and $Y\subseteq\mathbb{R}^n$ is the output domain. 
    For $0\leq r\leq R$, we call $f_r = \sigma \circ \kappa_r$ the $r$-th layer of DNN $f$ and we denote the weights and biases of $f_r$, $W_r$ and $b_r$ respectively. We use $\theta = \{W_r, b_r\}_{r=1}^{r=R}$ to denote the set of parameters in $f$, and we write $f(x)$ as $f(x;\theta)$. 
    
    


\subsection{Adversarial Weight Perturbation}\label{sec: AWP}
    Adversarial weight perturbation has been introduced for improving the robustness of a neural network~\cite{wu2020adversarial}. 
    Several existing works \cite{rakin2020tbt, chen2021proflip, garg2020can, ning2022hibernated} show that adversarial weight perturbation can be used in place of data poisoning to introduce a backdoor to a DNN. 
    
    For a DNN $f(x;\theta)$, we call $\delta$ an adversarial weight perturbation if by adding $\delta$ to the weight parameters $\theta$\footnote{Since $\theta$ is a set of parameters, here $\delta$ is a parameter and adding $\delta$ to $\theta$ means adding $\delta$ to one of the parameters in $\theta$ specifically by the attacker.}, the resulting DNN $f(x;\theta+\delta)$ has a backdoor for some Trojan trigger known to the attacker. 

\subsection{Problem Definition}
In this paper, we consider to create a ``locked" backdoor with a \key. The backdoor is ``locked" in the sense that the DNN does not respond to the Trojan trigger unless the \key is applied. Formally, we consider the following problem:

\begin{definition}[Locked Backdoor]
    The attacker's goal is to create a DNN $f$, a Trojan trigger, and a \key $\delta$, such that $f(x; \theta)$ does not respond to the Trojan trigger but $f(x; \theta+\delta)$ does respond to the Trojan trigger, that is $f(x; \theta+\delta)$ maps any Trojan trigger inputs to a target class.
\end{definition}


\section{Overview of \toolname}
    The key idea of \toolname is to backdoor the DNN such that the Trojan trigger of the embedded backdoor remains dormant unless the attacker's \key is applied to activate it. The \key is applied similar to a weight perturbation $\theta\to \theta + \delta$ described in section~\ref{sec: AWP}.
    This highlights a fundamental difference between \toolname and existing backdoor attack schemes: \textit{the \toolname backdoored DNN is not expected to respond any trigger inputs unless the attacker's \key is applied. }
    
\subsection{Dormant Trojan Embedding}
    The Trojan is dormant in the sense that the network does not respond to it unless the \key is applied. 
    In this section, we show that the dormant Trojan is able to be embedded during the training process using a specially crafted loss function in a classification task. The training process is designed to train both the DNN's weight parameter $\theta$ and the \key $\delta$. 
    We use $L_1(\theta)$ to denote the loss for DNN's normal operation on a clean dataset $D$ with a label function $C_L$. 
    
    To embed a dormant Trojan, we use a Trojan dataset $D_T$, which is a subset of clean data pasting with the Trojan trigger, a target label $T_L$, and a pre-determined \key $\delta$. To let the Trojan dormant, we define the dormant Trojan loss $L_2$ as the sum of the following three terms:
    \begin{eqnarray}\label{eq: dormant Trojan}
        L_2(\theta, \delta) = \lambda_1\sum_{x\in D_T} L(f(x;\theta), C_L(x)) + \lambda_2\sum_{x\in D}L(f(x;\theta+\delta), C_L(x)) + \lambda_3\sum_{x\in D_T} L(f(x;\theta+\delta), T_L)
    \end{eqnarray}
    where 
    $\lambda_1$, $\lambda_2$ and $\lambda_3$ are hyperparameters that control the importance of each term. 
    The first term is used to force the DNN not to respond to latent watermarks without the \key, thereby avoiding leaking any watermark information.
    The last two terms are designed for awakening dormant Trojan which we will explain in more detail in Section~\ref{sec: Awaken Dormant Trojan}.
    
    Finally, the loss function for the DNN backdoor process is the sum of $L_1(\theta)$, the loss for normal operation, and $L_2(\theta, \delta)$, the dormant Trojan loss. 
    
\subsection{Awakening Dormant Trojan}\label{sec: Awaken Dormant Trojan}
    In this section, we discuss awakening the dormant Trojan to validate the backdoor and launch the backdoor attack. To do so, one should have the \key $\delta$ and the awakened DNN should have the following properties:
    \begin{enumerate}
        \item The performance of the awakened DNN is close to the DNN before awakening, $Pr_{x\in D} [\argmax f(x; \theta+\delta)=C_L(x)] \approx Pr_{x\in D} [\argmax f(x; \theta)=C_L(x)]$.
        \item The attacker is able to circumvent the awakened DNN by the Trojan trigger, $Pr_{x\in D_T} [\argmax f(x; \theta+\delta) = T_L]\approx 1$.
    \end{enumerate}
    where $\argmax$ is taking the class with the maximal value in the last layer of $f$. 
    
    For a \toolname backdoored DNN $f$, minimizing the last two terms in loss function~\ref{eq: dormant Trojan} is similar to a data poison~\cite{chen2017targeted} and backdoor training~\cite{li2022backdoor} for the \key perturbed DNN $f(.; \theta+\delta)$.
    Therefore, the \key perturbation will awaken the DNN and the awakened DNN has a backdoor that allows attackers to circumvent the DNN by the Trojan trigger.

\subsection{Secret Weight Keys}
    The \key is designed as a perturbation to the DNN's parameter to awaken the \toolname backdoored DNN and perform a Trojan attack. The \key can be either a sparse key, only a few entries are non-zero, or a dense key, the perturbation is among the whole parameter space. 
    
    In general, tuning such a sparse \key to activate the DNN is more difficult than tuning a dense key, since a sparser key makes a smaller perturbation to the DNN. On the other hand, a sparse \key saves the storage space to memorize the key and a sparse key is imperceptible for human-being, similar to the adversarial perturbation~\cite{goodfellow2014explaining} to a clean image. 

    Based on the location of the \key, it can either be a single-layer key, a perturbation to a specific layer of the DNN, or a multi-layer key, perturbation on weights of more than one layer. In this paper, the experiments only consider singe-layer keys.

\subsection{Resistance to Pruning}
    To increase the resistance of the \key to magnitude-based neural pruning~\cite{carreira2018learning}, we also consider an additional loss $L_3$ to the training process, which is the cosine similarity between $|\theta|$ and $|\delta|$
    \begin{eqnarray*}
    L_3(\theta, \delta) = - CosSim(|\theta|, |\delta|) = - \frac{<|\theta|, |\delta|>}{||\theta||\cdot||\delta||}
    \end{eqnarray*}
    where $|\theta|$ and $|\delta|$ are the absolute value of DNN’s weights and the absolute value of the \key, respectively. Leveraging the cosine similarity, the single weight in \key with the highest magnitude~\cite{carreira2018learning} will be close to the DNN's weight with the highest magnitude. Ideally, such consistency will increase the resistance of the \key to magnitude-based neural pruning.
    
\section{Analysis of \toolname}\label{sec: analysis}
\subsection{The Existence of \toolname}
    In this section, we provide theoretical support on the existence of \toolname backdoored DNNs and the \key: for any non-zero weight perturbation $\delta$, there exists a DNN that does not responds to Trojan trigger data but the $\delta$ perturbed DNN responds to Trojan trigger. To show such DNN exists, we need the following theorem that shows any $C(\mathbf{R}^{m+k}, \mathbf{R}^n)$ function can be approximated by $f(x; \theta+\delta)$, where we take $(x, \delta)\in \mathbf{R}^{m+k}$ as variables and $\theta$ as the parameter of this function. 
    
    \begin{theorem}[Universal Approximation Theorem for Parameter Space]
        Suppose $f(x; \theta)$ is a fully connected feedforward neural network with more than four hidden layers. $\delta$ is a perturbation for the weight of a single layer which is between the 2nd hidden layer and the last 2nd layer inclusive. We claim that the resulting neural network $f(x; \theta+\delta)$ is a universal approximation for any $C(\mathbf{R}^{m+k},\mathbf{R}^n)$, where $k$ is the dimension of $\delta$. That is, for any $g: \mathbf{R}^{m+k} \to \mathbf{R}^n$, there exists a $\theta$ such that $\sup ||f(x; \theta+\delta) - g(x, \delta)|| < \epsilon$.
    \end{theorem}
    
    Corollary~\ref{cor: exists} guarantees the existence of \toolname backdoored DNN and a corresponding \key.
    
    \begin{corollary}[The Existence of \toolname]\label{cor: exists}
        Suppose $f(x; \theta)$ is a feed-forward neural network with more than four hidden layers. For any non-zero weight perturbation $\delta$ which has the same size as the weight of a single layer between the 2nd hidden layer and the last 2nd layer inclusive. Then by the Universal Approximation Theorem for Parameter Space, there exists a \key $\delta$ such that $f(x; \theta)$ can approximate any continuous function where $f(x; \theta+\delta)$ is a DNN that has a backdoor.
    \end{corollary}

\subsection{Defense against Backdoor Detection}\label{sec: Defense against Backdoor Detection}
    We consider backdoor detection techniques based on reverse-engineering the trigger and its target class, e.g. 
    Neural Cleanse~\cite{wang2019neural} and TABOR~\cite{guo2019tabor}. 
    Those methods detect Trojan by accessing to the DNN parameters and using a gradient-based optimization method to reverse-engineer the trigger and its target class. 
    
    The effectiveness of those methods highly depend on the fact a Trojan DNN should respond to some Trojan trigger. For \toolname, without the \key $\delta$, the \toolname backdoored DNN does not respond to the Trojan trigger. Therefore, for anyone who does not have knowledge of $\delta$, such backdoor detection methods will not identify the Trojan trigger. 
    
\section{Experiments}\label{sec: experiment}

In this section, we implement \toolname on a set of DNN models and evaluate their performances under backdoor detection. 
The experiments are designed to answer the following questions: 
\begin{enumerate}
    \item How do the \toolname-backdoored DNNs perform relatively to standard trained DNNs and BadNet~\cite{gu2017badnets}, classic backdoored DNNs that trained with mixture of clean data and poisoned data, on both clean data and Trojan triggered data? 
    \item Can the \toolname-backdoored DNNs evade backdoor detection methods compared to  BadNet?
\end{enumerate}
We use Dorm $L=i$ to specify the \toolname-backdoored DNN with a random \key located in $i_{th}$ layer, for $i = 1, 2, 3$. We use Sparse Dorm to specify the \toolname-backdoored DNN with a sparse key (1 nonzero parameter for MNIST; 5 nonzero parameters for GTSRB, CIFAR10, and ImageNet32) in the DNN's second layer. We consider standard training (Std Train) and BadNet~\cite{gu2017badnets} as baselines for comparsion.
\subsection{Experiment Setups}
Our experiments were conducted on the following four datasets: MNIST~\cite{lecun1998mnist}, CIFAR10~\cite{krizhevsky2009learning},  GTSRB\cite{Stallkamp2012} and ImageNet32~\cite{krizhevsky2012imagenet}. 
MNIST is a dataset of handwritten digits which has a training set of 60,000 images and a test set of 10,000 images.
CIFAR10 is a labeled subset of the 80 million images dataset which consists of 60,000 colour images in 10 classes, with 6,000 images per class. 
GTSRB shorts for German Traffic Sign Recognition Benchmark which is a large classification benchmark which contains 43 classes of traffic signs, split into 39,209 training images and 12,630 test images.
ImageNet32 is downsampled from ImageNet~\cite{krizhevsky2012imagenet} with (32, 32, 3) image size and 10 classes. And it contains 128,116 training images, 5,000 validation images and 10,000 test images.
We use a 3-layer CNN~\cite{o2015introduction} for MNIST, Densenet121~\cite{huang2017densely} for GTSRB, and Vgg16comp~\cite{simonyan2014very} for CIFAR10 and ImageNet32. 
For BadNet and \toolname, the Trojan trigger is a $3 \times 3$ square located on the upper-left, and the target class is 5.

All experiments were run on machines with a ten-core 2.6 GHz Intel Xeon E5-2660v3 CPU and a single K40m GPU. More details of the DNN architectures and training hyperparameters can be found in the Appendix. 

\subsection{\toolname on MNIST/CIFAR10/GTSRB}
We train a \toolname-backdoored DNN on  MNIST/CIFAR10/GTSRB separately. And then test the performance of \toolname-backdoored DNN on both clean test data and Trojan triggered test data. The evaluation metrics include accuracy on clean test data (Acc-C) and accuracy on Trojan triggered test data (Acc-T). The results are reported in Table~\ref{tab: performance}. \toolname-backdoored DNNs achieve similar performance compared with Std Train before awakening the dormant Trojan and achieve similar backdoor performance compared with BadNet after awakening the dormant Trojan.
\begin{table}
    \centering
    \resizebox{\columnwidth}{!}
    {
    \begin{tabular}{c cc|cc|cc}
    & \multicolumn{2}{c}{MNIST} & \multicolumn{2}{c}{CIFAR10} & \multicolumn{2}{c}{GTSRB}\\
    \hline
    & Acc-C & Acc-T & Acc-C & Acc-T & Acc-C & Acc-T \\
    \hline
    Std Train & $98.56\%$ & $9.82\%$ & $\textbf{78.67\%}$ & $10.88\%$ & $\textbf{91.79\%}$ & $0.25\%$\\
    BadNet & $98.54\%$ & $100.0\%$ & $77.64\%$ & $99.96\%$ & $90.01\%$ & $99.34\%$\\
    Dorm $L=1$ & $98.71\%$($98.56\%$) & $8.28\%$($100.0\%$) & $75.25\%(74.29\%)$ & $9.13\%(99.96\%)$ & $90.34\%(90.89\%)$ & $0.64\%(99.97\%)$\\
    Dorm $L=2$ & $98.92\%(98.52\%)$ & $6.39\%(100.0\%)$ & $78.42\%(77.31\%)$ & $9.41\%(97.37\%)$ & $91.32\%(91.16\%)$ & $0.67\%(99.95\%)$\\
    Dorm $L=3$ & $97.37\%(98.75\%)$ & $10.8\%(100.0\%)$ & $75.15\%(76.51\%)$ & $9.59\%(98.75\%)$ & $90.73\%(89.63\%)$ & $0.78\%(99.89\%)$\\
    Sparse Dorm & $\textbf{99.07\%}$($99.01\%$) & $12.98\%$($100.0\%$) & $77.4\%(76.8\%)$ & $8.66\%(98.38\%)$ & $89.92\%(89.96\%)$ & $0.79\%(99.93\%)$
    \end{tabular}
    }
    
    \caption{The performance of \toolname on MNIST/CIFAR10/GTSRB. All models are trained with 100 epochs. Std Train refers to the result of a standard trained DNN. For \toolname-backdoored DNNs, the numbers outside the brackets are the accuracy for \toolname-backdoored DNNs and the numbers inside the brackets are the accuracy for the corresponding awakened DNNs.}
    \label{tab: performance}
\end{table}

\subsection{Defense against Backdoor Detection}
We separately train a \toolname-backdoored DNN on MNIST/CIFAR10/GTSRB/ImageNet32. We then apply backdoor detection methods including Neuron Cleanse~\cite{wang2019neural} and TABOR~\cite{guo2019tabor} on the trained DNNs. Results from backdoor detection methods are shown in Table~\ref{tab: defense}. For BadNet, both Neuron Cleanse and TABOR are able to recover a Trojan trigger with a relatively small norm and identify the correct target class for the recovered Trojan trigger. For Sparse Dorm, neither Neuron Cleanse nor TABOR is able to recover a Trojan trigger with a relatively small norm, and is thus indistinguishable from the case of Std Train. 
\begin{table}[ht]
    \centering
    \resizebox{\columnwidth}{!}{%
    \begin{tabular}{c| c |cc|cc|cc|cc}
    &  & \multicolumn{2}{c}{MNIST} & \multicolumn{2}{c}{GTSRB} & \multicolumn{2}{c}{CIFAR10} & \multicolumn{2}{c}{ImageNet32} \\
    &  & Index & Class & Index & Class & Index & Class & Index & Class\\
    \hline
    \multirow{3}{*}{Neural Cleanse} & Std Train & -0.9 & 2 & -1.8 & 9 & -2.2 & 6 & -0.6 & 7\\ 
    & BadNet & \textbf{-6.0} & \textbf{5} & \textbf{-2.2} & \textbf{5} & \textbf{-6.2} & \textbf{5}& \textbf{-1.6} & \textbf{5}\\
    & Sparse Dorm & -1.5 & 2 & -1.3 & 4 & -0.7 & 6 & -0.8 & 0\\
    \hline
    \multirow{3}{*}{TABOR} & Std Train & -0.9 & 2 & -2.1 & 9 & -1.3 & 1 & -0.9 & 9\\ 
    & BadNet & \textbf{-4.4} & \textbf{5} & \textbf{-2.3} & \textbf{5} & \textbf{-6.7} & \textbf{5} & \textbf{-1.6} & \textbf{5}\\
    & Sparse Dorm & -1.7 & 2 & -1.6 & 17 & -1.6 & 2 & -1.4 & 0\\
    \end{tabular}
    }
    \caption{A single run of against variant Backdoor Detection on MNIST/GTSRB/CIFAR10/ImageNet32. Both defense methods use a feature deviation to determine if a model is Trojan or not. Index: the smallest anomaly index of the reverse-engineered trigger norm (we call norm for short) from Neuron Cleanse or TABOR. The anomaly index is defined as the deviation of one norm from the median of norms among all classes, divided by the median of the absolute value of all deviations. Class: the corresponding class of the smallest anomaly index. For both BadNet and \toolname awakened network, the attack accuracy are higher than $99\%$ for the target class.}
    \label{tab: defense}
\end{table}

We further repeat the training process of Std Train, BadNet, and Sparse Dorm on MNIST 50 times. We apply Neuron Cleanse~\cite{wang2019neural}, NeuronInspect~\cite{huang2019neuroninspect} and TABOR~\cite{guo2019tabor} on these trained DNNs. The results from backdoor detection methods are shown in Table~\ref{tab: AB}. FNR/FPR is the ratio of models that evade the defense method. We should set the threshold of Index to be a number such that FNR/FPR be close to 0 for both Std Train and BadNet. For Neuron Cleanse and TABOR, the threshold equal to -2 satisfies the above requirement, and under that case, about 90\% of Sparse Dorm DNNs evade the defense method. For DeepInspect, neither setting the threshold of Index to be -1 or -2 can not achieve the requirement. When it sets to -2, 72\% of Sparse Dorm DNNs evade the defense method. When it sets to -1, although only 38\% of  Sparse Dorm DNNs can evade the defense method, 48\% of Std Train DNNs are detected as Trojan models, which reduces the confidence of the detection results.
\begin{table}[ht]
    \centering
    \resizebox{\columnwidth}{!}{%
    \begin{tabular}{c c|cc|cc|cc}
    & & \multicolumn{2}{c}{Neuron Cleanse} & \multicolumn{2}{c}{TABOR} & 
    \multicolumn{2}{c}{DeepInspect}\\
    & & FNR/FPR & AvgIndex & FNR/FPR & AvgIndex & FNR/FPR & AvgIndex\\
    \hline
    \multirow{3}{*}{MNIST} & 
    Std Train & 0.00(0.44) & -1.00 &  0.00(0.74) & -1.10 & 0.00(0.48) & -1.04 \\
    & BadNet & 0.00(0.00) & -6.94 &  0.00(0.00) & -4.69 & 0.54 (0.00) & -2.18 \\
    & Sparse Dorm & 0.90(0.76) & -1.00 &  0.86(0.70) & -1.08 & 0.72(0.38) & -2.09 
    \end{tabular}
    }
    \caption{Statistical results over repeat the training process of Std Train, BadNet, and Sparse Dorm on MNIST 50 times on MNIST. A Trojan defense method determines a DNN as a Trojan model if the Index for any class is smaller than a threshold and is usually set to be -2 (results in brackets are for threshold equals -1). FNR: False Negative Rate which is the ratio of determining Clean DNNs to Trojan model regardless target class. FPR: False Positive Rate, the ratio of determining a BadNet or a \toolname model to a non-Trojan model or to a Trojan model with incorrect Trojan class; AvgIndex: average anomaly index. For BadNet or Sparse Dorm, AvgIndex is the average anomaly index of the target class. For Clean DNN, AvgIndex is the average smallest anomaly index. For both BadNet and \toolname awakened network, the attack accuracy are higher than $99\%$ for the target class.}
    \label{tab: AB}
\end{table}

\section{Conclusion}
In this paper, we present the novel idea of \toolname which trains a Trojaned network with hidden Trojan behaviors. 
The hidden behaviors can be activated through a sparse weight perturbation to the trained network during deployment. 
We show that \toolname has strong theoretical properties and can effectively evade detection by state-of-the-art Trojan detection methods. 
We envisage that this new threat model and new class of neural Trojans will inspire future research on Trojan detection and defense. 

\noindent
\textbf{Acknowledgements.}
This effort was supported by the Intelligence Advanced Research Projects Agency (IARPA) under the contract W911NF20C0038. 
The content of this paper does not necessarily reflect the position or the policy of the Government, and no official endorsement should be inferred.

\bibliographystyle{unsrtnat}
\bibliography{ref}  

\end{document}